\documentclass[pra,twocolumn,showpacs]{revtex4}
\usepackage{graphicx,amsfonts,amsmath,amssymb}
\usepackage{times}
\usepackage{txfonts}

\def\kbar{\protect\@kbar}
\def\@kbar{\relax \bgroup
\def\@tempa{\hbox{\raise.73\ht0
\hbox to0pt{\kern.25\wd0\vrule width.5\wd0
height.1pt depth.1pt\hss}\box0}}\mathchoice{\setbox0\hbox{$\displaystyle k$}
\@tempa}{\setbox0\hbox{$\textstyle k$}\@tempa}{\setbox0\hbox{$\scriptstyle k$}
\@tempa}{\setbox0\hbox{$\scriptscriptstyle k$}\@tempa}\egroup}

\begin{document}

\title{Quantum random walks using quantum accelerator modes}
\author{Z.-Y. Ma}
\affiliation{Clarendon Laboratory, Department of Physics,
University of Oxford, Parks Road, Oxford OX1 3PU, United Kingdom}
\author{M. B. d'Arcy}
\affiliation{Atomic Physics Division, National Institute of
Standards and Technology, Gaithersburg, Maryland 20899-8424, USA}
\altaffiliation[Also at ]{the Brookings Institution, 1775
Massachusetts Avenue NW, Washington, DC 20036-2103, USA}
\author{S. A. Gardiner}
\affiliation{Department of Physics, University of Durham,
Rochester Building, South Road, Durham DH1 3LE, United Kingdom}
\author{K. Burnett}
\affiliation{Clarendon Laboratory, Department of Physics,
University of Oxford, Parks Road, Oxford OX1 3PU, United Kingdom}

\date{\today}

\begin{abstract}
We discuss the use of high-order quantum accelerator modes to
achieve an atom optical realization of a biased quantum random
walk. We first discuss how one can create co-existent quantum
accelerator modes, and hence how momentum transfer that depends on
the atoms' internal state can be achieved. When combined with
microwave driving of the transition between the states, a new type
of atomic beam splitter results. This permits the realization
of a biased quantum random walk through quantum accelerator modes.
\end{abstract}

\maketitle

\section{Introduction}

Quantum accelerator modes are characterized by the efficient
transfer of large momenta to laser-cooled atoms by repeated
application of a spatially periodic potential
\cite{Oberthaler1999,Godun2000,dArcy2001}. Quantum accelerator
modes therefore constitute a potentially versatile technique for
manipulating the momentum distribution of cold and ultracold
atoms. Following the first observation of quantum accelerator
modes \cite{Oberthaler1999} there has been substantial progress in developing a
theoretical understanding of the mechanisms and structure that
underpin them \cite{Fishman2002,Bach2005}. This has permitted the
observation and categorization of higher-order quantum accelerator
modes \cite{Schlunk2003b}, demonstration that the momentum is
transferred coherently \cite{Schlunk2003a}, observation of the
sensitivity of the dynamics to a control parameter \cite{Ma2004},
and characterization of the mode structure in terms of number
theory \cite{Buchleitner2005}.

Quantum random walks have received attention due to their markedly
non-classical dynamics and their potential application as
search algorithms in practical realizations of quantum
information processors  \cite{Aharanov1993,Bach2002}. In this paper, we report an investigation
into the use of high-order quantum accelerator modes to implement
a quantum random walk in the momentum space distribution of cold
atoms \cite{Buerschaper2004}. This method is more robust and easier
compared with other recent proposals for implementing quantum
random walks using ion traps \cite{Travaglione2002}, microwave or optical
cavities \cite{Sanders2002} and optical lattices \cite{Dur2002},
and should make feasible quantum random walks of a few hundred
steps. This would be a useful experimental tool for information
processing.

In this paper we first survey the experimental phenomenology and
theoretical understanding of quantum accelerator modes. We then
discuss how the generation of specific quantum accelerator modes
can be experimentally controlled. Based on these techniques, we
explain how internal-state-dependent momentum transfer can be
achieved, which will permit coherent beam-splitting. Finally, we
show how this could be applied to realize experimentally a biased
quantum random walk procedure.

\section{Overview of quantum accelerator modes}

\subsection{Observation of atomic quantum accelerator modes}

Quantum accelerator modes are observed in the $\delta$-kicked
accelerator system \cite{dArcy2001}. In the atom optical
realization of this system, a pulsed, vertical standing wave of
laser light is applied to a cloud of laser-cooled atoms
\cite{Oberthaler1999,Godun2000,dArcy2001,Fishman2002,Bach2005,Schlunk2003a,Schlunk2003b,Ma2004}.
The corresponding Hamiltonian can be written:
\begin{equation}
\hat{H}=\frac{\hat{p}^{2}}{2m}+mg\hat{z}-\hbar \phi _{d}[1+\cos
(G\hat{z} )]\sum_{n}\delta (t-nT), \label{Eq:HCOM}
\end{equation}
where $\hat{z}$ is the vertical position, $\hat{p}$ is the
momentum, $m$ is the atomic mass, $g$ is the gravitational
acceleration, $t$ the time, $T$ the kicking pulse period,
$G=2\pi/\lambda _{\mbox{\scriptsize spat}}$, where $\lambda
_{\mbox{\scriptsize spat}}$ is the spatial period of the potential
applied to the atoms, and $\phi_{d}$ quantifies the kicking
strength of laser pulses, i.e. the laser intensity. This
Hamiltonian is identical to that of the $\delta$-kicked rotor, as
studied experimentally by the groups in Austin \cite{Moore1995},
Auckland \cite{Ammann1998}, Lille \cite{Szriftgiser2002}, Otago
\cite{Duffy2004}, London \cite{Jones2004}, and Gaithersburg
\cite{Ryu2005}, apart from the addition of the linear
gravitational potential $mg\hat{z}$; this linear potential is critical
to the generation of the quantum accelerator modes.

In the experiments performed to date to observe quantum
accelerator modes, cesium atoms are trapped and cooled in a
magneto-optic trap to a temperature of $5\mu$K. They are then
released from the trap and, while they fall, a series of standing
wave pulses is applied to them. Following the pulse sequence, the
momentum distribution of the atoms is measured by a time of flight
method, in which the absorption of the atoms from a sheet of
on-resonant light through which they fall is measured. The quantum
accelerator modes are characterized by the efficient transfer of
momentum, linear with pulse number, to a significant fraction
($\sim 20$\%) of the atomic ensemble. 

The spatially periodic potential experienced by the atoms in the
far off-resonant standing light wave is due to the ac Stark shift.
We can therefore write $\phi_{d}=\Omega_{R}^{2}t_{p}/8\delta _{L}$
\cite{Godun2000}, where $\Omega_{R}$ is the Rabi frequency at the
intensity maxima of the standing wave, $t_{p}$ is the pulse
duration and $\delta_{L}$ is the red-detuning of the laser
frequency from the $6^{2}S_{1/2}\rightarrow 6^{2}P_{1/2}$,
($|F=4\rangle \rightarrow |F'=3\rangle$) D1 transition of cesium.
In these experiments, the standing wave light is produced by a
Ti:Sapphire laser; the maximum intensity of the laser beam is
$\sim 1\times 10^{4}$ mW cm$^{-2}$. Within the regime where
spontaneous emission can be ignored \cite{Godun2000}, the detuning
can be modified over a range of order $30$\thinspace GHz, so that
the kicking strength can be changed by roughly an order of
magnitude. If $\delta _{L}=2\pi \times 30$\thinspace s$^{-1}$,
$\phi_{d} \simeq 0.8\pi$.

Quantum accelerator modes may be observed in $\delta$-kicked
accelerator dynamics when $T$ is close to values at which
low-order quantum resonances occur in the quantum $ \delta
$-kicked rotor, i.e. integer multiples of the half-Talbot time
$T_{1/2} = 2\pi m/\hbar G^{2}$ (so named because of a similarity
of this quantum resonance phenomenon to the Talbot effect in
classical optics). In the case of the Oxford experiment, $T_{1/2}
= 66.7\mu$s \cite{Goodman1996}.

\subsection{$\epsilon$-classical theory and high-order modes}

In 2002 Fishman, Guarneri, and Rebuzzini (FGR) \cite{Fishman2002}
used an innovative analysis, termed the $\epsilon$-classical
expansion, to explain the occurrence and characteristics of the
observed quantum accelerator modes. This theoretical framework
predicted the existence of higher-order modes, which was
subsequently verified experimentally \cite{Schlunk2003b}. Our
later discussion focuses on these higher-order modes, so we
briefly summarize the $\epsilon$-classical theory here.

In the $\delta$-kicked rotor, the spatial periodicity of the
kicking potential means that momentum is imparted in integer
multiples of $\hbar G$. This spatial periodicity also means that
the dynamics of any initial atomic momentum state are equivalent
to those of a state in the first Brillouin zone, $0 \leq p < \hbar
G$, i.e., the momentum modulo $\hbar G$. This is the
quasimomentum, and hence it is a conserved quantity in the kicking
process \cite{Ashcroft1976}. 

The presence of gravitational
acceleration in the $\delta$-kicked accelerator breaks this
periodic translational symmetry. Transforming to a freely-falling
frame removes the $mgz$ term from the Hamiltonian; consequently
quasimomentum conservation is observed, {\it in the freely-falling
frame}. Conservation of quasimomentum means that different
quasimomentum subspaces evolve independently. The FGR theory makes
use of this property to decompose the system into an assembly of
``$\beta$-rotors'' \cite{Fishman2002,Bach2005}, where the
quasimomentum $=\beta\hbar G$ and $\beta \in [0,1)$.

The linear potential due to gravity makes its presence felt by
changing, relative to the case of the $\delta$-kicked rotor, the
phase accumulated over the free evolution between kicks. This
means that quantum resonance phenomena different  from those
observed in the $\delta$-kicked rotor occur. For values of $T$
close to $\ell T_{1/2}$, where $\ell \in \mathbb{Z}$, certain
states permit rephasing, and hence, within any given quasimomentum
subspace, the projection of the initial condition onto states
which are appropriately localized within (periodic) position {\it
and \/} momentum space are coherently accelerated away from the
background atomic cloud. The momentum of the accelerated population
increases linearly with the number of kicks, and it is this which
constitutes a quantum accelerator mode.

The closeness of the kicking period $T$ to integer multiples of
the half-Talbot time $T_{1/2}$ is formalized in the FGR theory by
the smallness parameter $\epsilon=2\pi (T/T_{1/2}-\ell)$. In the
limit of $\epsilon \rightarrow 0$ it is possible to simplify the
dynamics of the operator-valued observables to a set of effective
classical (or {\it pseudoclassical}) mapping equations, separate
but identical for each independently evolving quasimomentum
subspace, or $\beta$-rotor. If we define the parameters $K=\phi_{d}|\epsilon |$ and
$\Omega = gGT^{2}/2\pi$, these mapping equations can be written:
\begin{subequations}
\begin{gather}
J_{n+1}=J_{n}-\mathrm{sgn}(\epsilon)2\pi \Omega -K\sin \theta
_{n},
\label{eq:mapping1}\\
\theta_{n+1}=\theta_{n}+\mathrm{sgn}(\epsilon)J_{n+1}\mathrm{mod}(2\pi),
\label{eq:mapping2}
\end{gather}
\label{eq:mapping}
\end{subequations}
where $J_{n}$ and $\theta _{n}$ are the transformed momentum and
position variables, respectively, just before the $n$th kick. A
quantum accelerator mode corresponds to a stable island system,
centered on a periodic orbit, in the stroboscopic phase space
generated by the mapping of Eq.\ (\ref{eq:mapping}). As the
dynamics of interest take place within stable islands and are
therefore approximately harmonic, the usefulness of this
pseudoclassical picture actually extends over a broader range of
$\epsilon$ than might otherwise be expected \cite{Bach2005}.

A given island system is specified by the pair of numbers
$\mathfrak{p}$, the order of the fixed point, and $\mathfrak{j}$,
the jumping index, and the quantum accelerator mode can be
likewise classified. Physically, $\mathfrak{p}$ is the number of
pulse periods a ``particle'' initially on a periodic orbit takes
before cycling back to the initial point in the reduced
phase-space cell, while $\mathfrak{j}$ is the number of unit cells
of extended phase space traversed by this particle in the momentum
direction per cycle, i.e., $J_{n\mathfrak{p}}= J_{0}+2\pi n
\mathfrak{j}$. Transforming back to the conventional linear
momentum in the accelerating frame, after $N$ kicks, the momentum
of the accelerated atoms is given by:
\begin{equation}
p_{N} \simeq 2\pi N\left[ \frac{\mathfrak{j}}{\mathfrak{p}}
+\mbox{sgn}(\epsilon)\Omega \right] \frac{\hbar G }{|\epsilon |}.
\label{eq:moment}
\end{equation}

The first quantum accelerator modes to be observed were those for
which $\mathfrak{p}=1$ \cite{Oberthaler1999}. Since then, others
with orders as high as $23$ have been observed
\cite{Schlunk2003b}. We shall now focus on these higher-order
modes.

\subsection{Coexistence of quantum accelerator modes}

The phase space generated by application of the mappings of
Eq.\ (\ref{eq:mapping}) changes as the parameters $K$ and $\Omega$
are varied. In experiments to date, $K$ and $\Omega$ have
generally been varied simultaneously by scanning $T$, and hence also
$\epsilon$ \cite{Buchleitner2005}. The structure of phase space
may also be altered by varying $\phi_{d}$, and hence $K$ alone, or
by varying $g$, and hence $\Omega$ alone. As the phase space
changes, two or more distinct island chains, specified by
different $(\mathfrak{p},\mathfrak{j})$, can coexist. This means
that the corresponding quantum accelerator modes can be
simultaneously produced by kicking the atoms. Hence different
amounts of momentum can be transferred to several classes of the
atoms evolving from the initial ensemble. This phenomenon may
offer the possibility of building an atomic beam splitter. As we
shall argue below, it may also permit a quantum random walk to be
realized in the atomic sample. We shall now consider some examples
of the effect of altering $\phi_{d}$ and $g$.

\subsection{Tuning the kicking strength}

\begin{figure}[tbp]
\centering
\includegraphics[width=9cm]{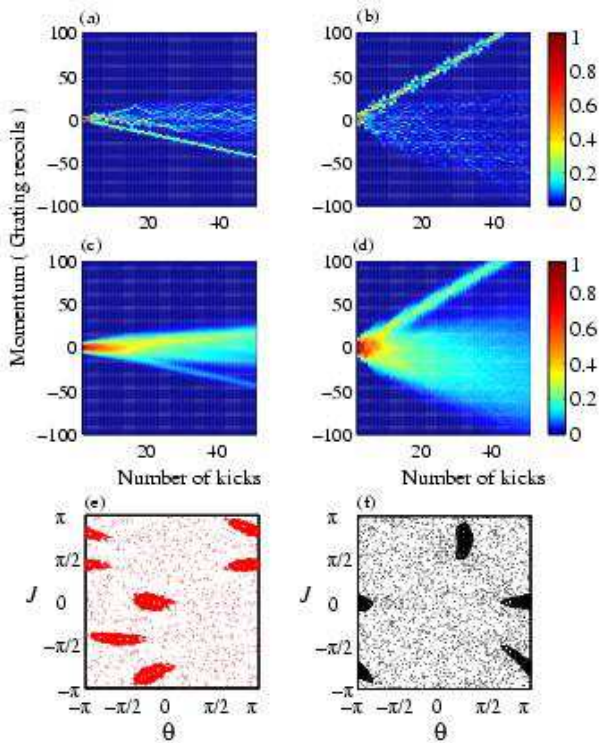}
\caption{(Color online) Numerical simulation for the quantum accelerator modes
that are produced with $g=9.81$\thinspace ms$^{-2}$,
$T=132.0\mu$s. In column 1 (a, c, e), $\phi_{d}=0.8\pi$ and the
$(5,2)$ quantum accelerator mode is produced; in column 2 (b, d,
f), $\phi_{d}=2.4\pi$ and the $(3,1)$ quantum
accelerator mode results. (a) and (b) show the momentum variation
with the number of kicks for a single initial plane wave with
$\beta =0$, (c) and (d) show the evolution of an atomic cloud with
initial temperature $5\mu$K, and (e) and (f) show stroboscopic
Poincar\'{e} sections determined by Eq.\ (\ref{eq:mapping}), with
$T=132.0\mu$s ($\epsilon =-0.135$). The colorbar indicates the
population, in arbitrary units.} \label{fig:qwsg09}
\end{figure}

We first examine the high-order quantum accelerator mode close to
the Talbot time, $T_{\mbox{\scriptsize T}}=133.4\mu$s, for the
case of a single initial quasimomentum state. We take the value
$\beta=0$, and consider the case where $T=132.0\mu$s and the local
gravity value $g=9.81$\thinspace ms$^{-2}$. We apply two different
kicking strengths to the atoms, $\phi_d=0.8\pi $ and $2.4\pi$. The
results of our numerical simulations, shown in
Figs.\ \ref{fig:qwsg09}(a) and \ref{fig:qwsg09}(b), demonstrate
that atoms evolving under $\phi_d=0.8\pi$ undergo a negative
momentum transfer, while the atoms experiencing $\phi_d=2.4\pi$
undergo a positive momentum transfer. The phase maps given in
Figs.\ \ref{fig:qwsg09}(e) and \ref{fig:qwsg09}(f), along with
Eq.\ (\ref{eq:moment}), show that for the lower kicking strength
the quantum accelerator mode is $(5,2)$, while for the higher
kicking strength the quantum accelerator mode is $(3,1)$.

Within a given quasimomentum subspace, the values of $J$ available
for the initial state are equal to $(k+\beta)|\epsilon|$, where $k$ is an
integer. In the
case of a narrow initial momentum distribution, we expect the
value of $\beta$, offsetting the available momentum spectrum, to
affect the significance of that subspace's contribution to a
physically observable quantum accelerator mode. Such an effect is
clearly of decreasing relevance as $\epsilon$ vanishes
\cite{Fishman2002,Bach2005}. This general observation is borne out
by numerical simulation.
\begin{figure}[tbp]
\centering
\includegraphics[width=6cm]{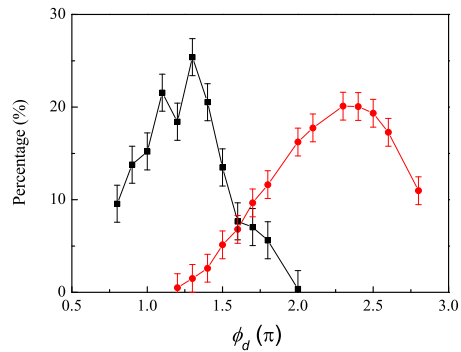}
\caption{(Color online) Variation of the percentage of atoms in the $(5,2)$ and
$(3,1)$ quantum accelerator modes as $\phi_{d}$ changes. The
atomic ensemble is prepared at $5\mu$K, and $T=132.0\mu$s. The
$(5,2)$ mode (squares) appears with lower kicking strength, and
$(3,1)$ mode (circles) appears when the kicking strength
increases. The error bar illustrates the typical spread of
population in a specific quantum accelerator mode obtained from
the simulation.} \label{fig:sumover}
\end{figure}

For the case of a thermal atomic cloud, such as the $1\times
10^{7}$ atoms at $5\mu$K with a Gaussian initial momentum
distribution in which all $\beta$ are
populated more-or-less equally, as used in our experiments, the dependence of the
acceleration on the kicking strength is shown in
Figs.\ \ref{fig:qwsg09}(c) and \ref{fig:qwsg09}(d). As expected,
the $(5,2)$ and $(3,1)$ quantum accelerator modes, respectively,
are produced. For this system, we can ask at which kicking
strength the different quantum accelerator modes appear. The
variation of the population in each quantum accelerator mode as a
function of $\phi_d$, deduced from the numerical simulations, is
shown in Fig.\ \ref{fig:sumover}. When the kicking strength is less
than $1.2\pi$, the atoms occupy the $(5,2)$ mode and the $(3,1)$
mode is absent. As one increases the kicking strength, the $(5,2)$
mode gradually disappears while the $(3,1)$ mode comes to
dominate; on further increasing $\phi_{d}$, the $(5,2)$ mode dies
completely. There is a range of $\phi_{d}$, centered on the value
$1.6\pi$, where the quantum accelerator modes co-exist and atoms
can be accelerated in two different modes simultaneously, with
different directions of momentum transfer.

\subsection{Tuning the effective gravitational acceleration}

\begin{figure}[tbp]
\centering
\includegraphics[width=9cm]{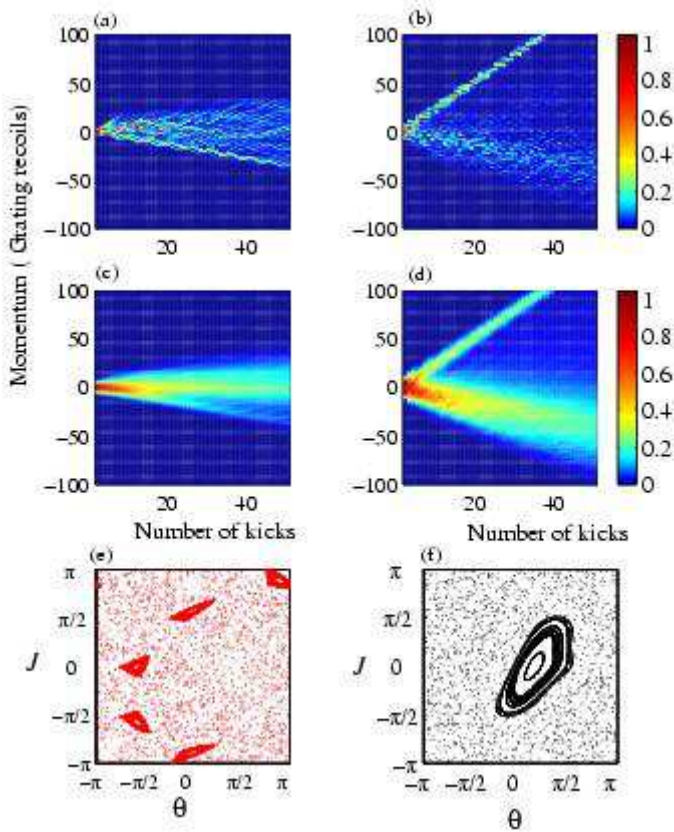}
\caption{(Color online) Numerical simulation for the quantum accelerator modes
that are produced with $g=20.10$\thinspace ms$^{-2}$,
$T=137.0\mu$s. In column 1 (a, c, e), $\phi_{d}=0.8\pi$ and the
$(5,-4)$ quantum accelerator mode is produced; in column 2 (b, d,
e), $\phi_{d}=2.4\pi$ and the $(1,-1)$ (b, d, f) quantum
accelerator mode results. (a) and (b) show the momentum variation
with the number of kicks for a single initial plane wave with
$\beta =0$, (c) and (d) show the evolution of an atomic cloud with
initial temperature $5\mu$K, and (e) and (f) show stroboscopic
Poincar\'{e} sections determined by Eq.\ (\ref{eq:mapping}), with
$T=137.0\mu$s ($\epsilon =0.336$). The colorbar indicates the
population, in arbitrary units.} \label{fig:qwsg10}
\end{figure}

It is possible to vary the value of the effective gravitational
acceleration applied to the atoms in our experiment, and hence
$\Omega$. This is accomplished by using an electro-optic modulator
to vary the phase difference between the down-going and
retro-reflected beams, and hence to move the profile of the
standing wave \cite{dArcy2001,Ma2004}. This allows us to reach
other parameter combinations that yield simultaneous acceleration
in different directions. For example, if we tune the effective
gravity to $20.10$\thinspace ms$^{-2}$ and choose a kicking period
of $T=137.0\mu$s, the occupied quantum accelerator mode is
$(5,-4)$ for the atoms which experience $\phi_{d}=0.8\pi$ and
$(1,-1)$ for those which evolve under $\phi_{d}=2.4\pi$. The
results of the corresponding numerical simulations are shown in
Fig.\ \ref{fig:qwsg10}.

Hence the momentum transferred by each kick can be varied by
properly selecting the effective gravitational acceleration,
kicking period and kicking strength in order to single out
particular quantum accelerator modes. We have also found a large
number of other conditions where atoms are accelerated in different quantum
accelerator modes, according to the value of $\phi_{d}$.

\section{Incorporation of electronic degrees of freedom}

\subsection{Using an electronic superposition
state}

Within a given parameter regime, i.e., for particular 
values of $\phi_{d}$ and $g$, and restricting ourselves to a single plane-wave
as the initial condition, it is not possible to  optimally
occupy two quantum accelerator modes for simultaneous
acceleration. This can be understood by realizing that coexisting quantum
accelerator modes must necessarily occupy different regions of pseudoclassical
phase space.

An efficient way to obtain simultaneous momentum
transfer in two directions is to start with a coherent
superposition of internal atomic states so as to optimally change
$\phi_d$ separately. These internal states, produced using a
microwave pulse, experience different kicking strengths.
This
allows us to have a situation where the same initial motional state
experiences two different two different kicking strengths, and maximally
occupies two different quantum accelerator modes, resulting in different
momentum transfers to the two parts of the superposition.

Considering two general electronic states $|a\rangle$ and
$|b\rangle$, the desired model Hamiltonian has the form
\cite{Schlunk2003a}
\begin{equation}
\hat{H}_{ab} = \hat{H}(\phi_{d}^{a})|a\rangle\langle a| +
\hat{H}(\phi_{d}^{b})|b\rangle\langle b|
+\frac{\hbar\omega_{ab}}{2}(|b\rangle\langle b|-|a\rangle\langle
a|),
\end{equation}
where $\hbar\omega_{ab}$ is the energy gap between $|a\rangle$ and
$|b\rangle$, and $\hat{H}(\phi_{d}^{a})$ and
$\hat{H}(\phi_{d}^{b})$ are equal to the atomic center of mass
Hamiltonian of Eq.\ (\ref{Eq:HCOM}), with $\phi_{d}=\phi_{d}^{a}$
and $\phi_{d}^{b}$, respectively. In our experiments, $|a\rangle$
may correspond the $|F=3,m_F=0\rangle$ substate of the ground
state of cesium, and $|b\rangle$ may correspond to the
$|F=4,m_F=0\rangle$ substate; henceforth these substates will be
denoted $|a\rangle$ and $|b\rangle$, respectively.

\subsection{Use of microwave pulses for state preparation}

The population of cesium atoms in the states $|a\rangle$ and
$|b\rangle$ can be modified by a $9.18$\thinspace GHz microwave
pulse, resonant with the $|b\rangle \rightarrow |a\rangle$
hyperfine transition \cite{Schlunk2003a}. The $9.18$\thinspace GHz
difference between the transition frequencies from the states
$|a\rangle$ and $|b\rangle$ to any given excited state means that
atoms in the two internal states will experience different values
of $\phi_{d}$ when exposed to laser light of a particular
intensity and detuning. 

A coherent superposition of $|a\rangle$
and $|b\rangle$ can be achieved experimentally by applying a
$\pi/2$ microwave pulse to a sample of atoms in state $|b\rangle$,
in which they are trapped and cooled. The intensity and detuning
of the light creating the kicking potential can be selected so as
to apply the correct values of $\phi_{d}$ to the states
$|a\rangle$ and $ |b\rangle$ to permit efficient population of the
required quantum accelerator modes. For example, with our current
experimental setup, it is feasible to have a value
$\phi_{d}=0.8\pi $ for state $|a\rangle $, while the corresponding
value for state $|b\rangle $ is $\phi_{d}=2.4\pi$. Without any
alteration to the effective value of $g$, atoms in $|b\rangle $
state will be kicked in one direction [in the $(3,1)$ quantum
accelerator mode] while atoms in $|a\rangle $ state will be kicked
in the other [in the $(5,2)$ quantum accelerator mode], as shown
in Fig.\ref{fig:qwsg09}. The transfer of momentum is therefore
dependent on the internal state, which is just what one needs for
a beam splitter. This may well lead to a new type of
interferometry based on this beam splitting mechanism and will be
the subject of future investigations.

\subsection{State-dependent evolution}

\begin{figure}[tbp]
\centering
\includegraphics[width=9cm]{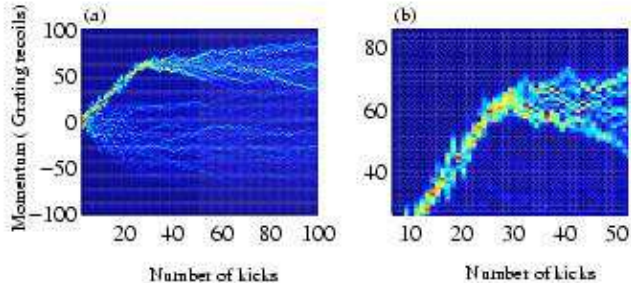}
\caption{(Color online)(a) Momentum variation with number of kicks, for $\beta
=0$, $\phi_{d}=2.2\pi$ for state $|a\rangle $ and $\phi_{d}=0.7\pi
$ for state $|b\rangle$, with $T=132.0\mu$s and $g=9.81$\thinspace
ms$^{-2}$. A state-flipping microwave pulse is applied after the
$25$th kick. The zoom-in around the switch point is shown in (b).}
\label{fig:evidenta}
\end{figure}

In this paper, however, we are focusing on the application of the
technique of simultaneous momentum transfer that quantum
accelerator modes provide to quantum random walks. The
state-dependence of the momentum transfer permits the
state-dependent evolution required for a quantum, rather than
classical, random walk. With atoms initially in a superposition of
the $|a\rangle $ and $|b\rangle $ states, we can apply kicks to
accelerate the atoms in the two states in different directions.

To investigate how the methods of manipulating the internal state
of the atoms permit momentum control, we numerically simulate a
sequence in which we accelerate atoms in state $|b\rangle$ for
$25$ kicks with $\phi _{d}=2.2\pi$, and we then apply a $\pi$
microwave pulse to pump all atoms from state $|b\rangle$ into
state $|a\rangle$, for which $\phi_{d}=0.7\pi$. $T=132.0\mu$s and
$g=9.81$\thinspace ms$^{-2}$ are kept constant during the process.
The results of the simulation are shown in Fig.\
\ref{fig:evidenta}. After the switch, atoms in $|b\rangle $ cease
increasing momentum in their original direction and about $30$\%
of them begin to accumulate momentum in the opposite direction,
corresponding to the quantum accelerator mode with the lower
kicking strength. Optimization of the efficiency of transfer from
one quantum accelerator mode to the other needs a more detailed
investigation, as we now discuss.

\subsection{Optimizing the switch property}

An ideal switch between different momentum transfer modes requires
the wavefunction of one quantum accelerator mode to have an
overlap with the other mode at the time of switching. From the FGR
analysis, this implies that better switching efficiency will occur
when the stable islands in pseudoclassical phase space for the
two quantum accelerator modes overlap \cite{Buchleitner2005}. 

\begin{figure}[tbp]
\centering
\includegraphics[width=8.5cm]{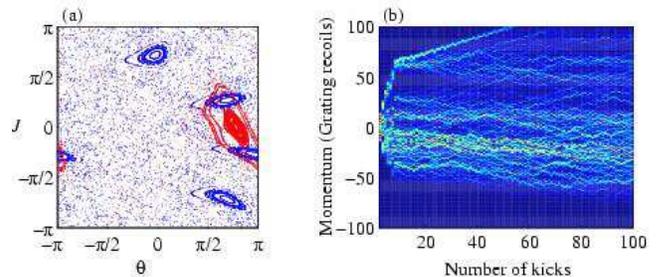}
\caption{(Color online) (a) Phase map of quantum accelerator modes, with $\beta
=0.1$, $T=131.0\mu$s, $g=7.26$\thinspace ms$^{-2}$,
$\phi_{d}=3.8\pi$ for state $|a\rangle$ (red dots, ) and
$\phi_{d}=0.6\pi $ for state $|b\rangle $ (blue dots). (b)
Momentum variation with number of kicks; a state-flipping $\pi$
pulse occurs after the $8$th kick.} \label{fig:overlap}
\end{figure}

This
is illustrated in Fig.\ \ref{fig:overlap}, where
$g=7.26$\thinspace ms$^{-2}$, $T=131.0\mu$s, $\phi_{d}=0.6\pi$ and
$3.8\pi$ for two different states. The overlap between the
stable islands for the lower kicking strength [mode $(1,0)$,
blue dots] and the higher kicking strength [mode $(4,1)$, red
dots] in Fig.\ \ref{fig:overlap}(a) is greater than that in Figs.\
\ref{fig:qwsg09}(e) and \ref{fig:qwsg09}(f), or Figs.\
\ref{fig:qwsg10}(e) and \ref{fig:qwsg10}(f). This, as we would
expect, leads to a more efficient transfer of population between
the quantum accelerator modes when the atomic internal state is
flipped by a microwave pulse, as shown by the comparison between
Fig.\ \ref{fig:overlap}(b) and Fig.\ \ref{fig:evidenta}(b). About
$80$\% of the atoms are successfully transferred from one mode to
the other. 

The $\epsilon$-classical map thus provides the capability of
using the overlap criterion to search in parameter space to find
the best switching condition. A complete search of the relevant
phase space is a substantial enterprise, and will be part of a
longer term effort to optimize the operation of a practical
random-walker.

\section{Near-ideal biased quantum random walk}

We now turn to the implementation of a quantum random walk using
the state-dependent acceleration process we have just described.
Applying a $\pi /2$ microwave pulse after each kick is equivalent
to the \textquotedblleft coin-flipping" process introduced by
Aharanov in his discussion of a quantum random walk
\cite{Aharanov1993}. In this section, we would like to show how we
could use quantum accelerator modes to implement a quantum random
walk. 

This scheme also introduces different features from the
Aharanov model, and we therefore name this model a ``biased''
quantum random walk in momentum space. In a biased quantum walk,
the ``coined'' state, which determines the direction atoms move
in by the extra degree of freedom of ``sides'' (discussed in
Ref.\ \cite{Aharanov1993}),
is the pair of hyperfine states of the atoms and the momentum
transfer per step, i.e., the walk speed, is determined by the
order of quantum accelerator modes. This can be altered [see Eq.\
(\ref{eq:moment})] by selecting different values of the parameters
$K$ and $\Omega$ that determine the acceleration. In this way
atoms can be made to perform a Hadamard-style quantum random walk
in momentum space. 

It is important to note that atoms are divided
to three different classes in the case of quantum accelerator
modes considered here: two of them fall into two different accelerator modes, thus
obtaining different momentum changes in each step, and the rest of
the atoms are ``left behind''. There is an overall recoil in the
opposite direction to the quantum accelerator modes
\cite{NoteRecoil}, but within this the motion is diffusive rather
than the coherent motion of quantum accelerator modes. In order to
understand better how such a system could be used to realise a
quantum random walk we propose the following simplified model. Our
model is a ``biased quantum random walk'' for our coined quantum
accelerator mode, where atoms not only walk in two different
directions, but can be left behind. 

The walk operator $S$ then
reads,
\begin{equation}
\begin{split}
S =&(1-\gamma )(|a \rangle \langle a |\otimes
\sum_{i}|i-\delta _{1}\rangle \langle i|  \\
&+|b \rangle \langle b |\otimes \sum_{i}|i+\delta _{2}\rangle
\langle i|)+\gamma \sum_i |i\rangle \langle i|,
\end{split}
\label{eq:newsoperator}
\end{equation}
where integers $i$ indicate the momentum states, and $\delta _{1}$
and $\delta _{2}$ corresponding to selected accelerator modes of
$(\mathfrak{p}_{1},\mathfrak{j}_{1})$ and
$(\mathfrak{p}_{2},\mathfrak{j}_{2})$. Here $\gamma$ is the
``leaving behind'' amplitude.

\begin{figure}[tbp]
\centering
\includegraphics[width=9cm]{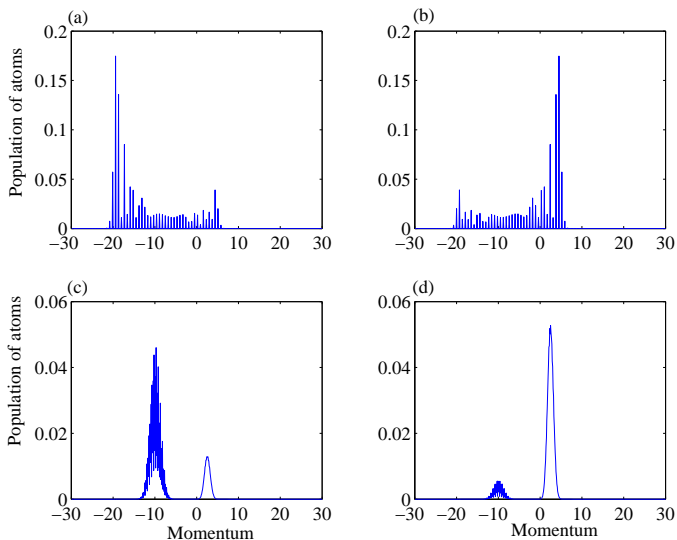}
\caption{The momentum distribution (in arbitrary units) of the
biased quantum random walk with a Hadamard coin after 50 steps,
starting in state $|a\rangle \otimes |0\rangle$ for (a) and (c),
and state $|b\rangle \otimes |0\rangle $ for (b) and (d). The
parameter $\gamma$, defined in Eq.\ \ref{eq:newsoperator}, is $0$
for (a, b) and $0.5$ for (c, d). The momentum increase is $0.25$
units per step to the negative direction and $0.1$ units per step
to the positive direction.} \label{fig:qrw1608varyinitial}
\end{figure}

The results of the numerical simulation of this biased quantum
random walk are shown in Fig.\ \ref{fig:qrw1608varyinitial}.
Quantum accelerator modes increase the momentum of a group of
atoms linearly with the number of kicks, and this means that the
effective ``diffusion'' of the biased walk will also be linearly
proportional to the number of kicks, or ``superdiffusive.'' 
We should expect atoms moving
faster in one direction than the other due to the difference in
the walking speeds of the two occupied quantum accelerator modes.
Walks with non-zero values of the parameter $\gamma$ have very
different distributions from those with $\gamma=0$. In particular
walks with $\gamma\neq 0$ will fill up the momentum gaps produced
by a ``pure'' $\gamma =0$ quantum random walk.

From Fig.\ \ref{fig:overlap} about $80$\% of atoms have a good
switch from one mode to another and $20$\% are left behind,
for appropriate values of $\beta$, $g$, and $\phi_{d}$. In
this way, atoms could perform quantum random walk for several
steps. A future study to perfect the switching property is
necessary. The value of such walks in search algorithms, and ways
of varying $\gamma$, will be the subject of future work. In this
paper we simply want to emphasis the potential interest and value
of state-dependent momentum transfer in quantum accelerator modes,
of the type we investigate here.

\section{Conclusions}

In conclusion, we have described a novel way to produce
state-dependent momentum transfer in a group of atoms. We believe
that this offers a new route to produce quantum random walks in
the laboratory with feasible experimental parameters. In
particular, the next generation of experiments with enhanced
velocity selection will put practical realizations well within
reach. The state-dependent walk controlled through the parameters
of the external perturbation is worthy of investigation in its own
right. 
There are three independent control parameters in the basic $\delta$-kicked
accelerator, namely the
driving strength, the effective gravitational acceleration, and the value of the
commutator $|\epsilon|$. In an atom-optical configuration these can all be tuned
independently.  There are thus many parameter regimes available 
particularly when considering the additional degrees of freedom offered by
superposition states.  The full range of such phenomena, and their relevance to
quantum random walks, quantum resonances and quantum chaos in
superposition states, awaits exploration.

\section*{Acknowledgements}

We thank R. M. Godun, S. Fishman, I. Guarneri, L. Rebuzzini, and
G. S. Summy. We acknowledge support from the UK EPSRC, the Royal
Society, and the Lindemann Trust.


\begin{thebibliography}{99}
\bibitem{NoteRecoil}
If all possible initial conditions are populated, i.e., the
initial distribution of atoms is large enough to cover several
phase-space cells, then in the freely-falling frame the average
momentum is conserved. Thus, the unaccelerated cloud always on
average ``recoils'' from the quantum accelerator modes.
\end{thebibliography}
\end{document}